\documentclass{appolb}
\usepackage{graphicx,amsmath,amssymb}

\begin{document}
\title{Collectivity in ultra-peripheral heavy-ion\\ and e+A collisions%
\thanks{Presented at ``Diffraction and Low-$x$ 2024'', Trabia (Palermo, Italy), September 8-14, 2024.}  
}
\author{Bjoern Schenke
\address{Physics Department, Brookhaven National Laboratory, Upton, NY 11973, USA}
\\[3mm]
{Chun Shen
\address{Department of Physics and Astronomy, Wayne State University, Detroit, Michigan 48201, USA}
\address{RIKEN BNL Research Center, Brookhaven National Laboratory, Upton, NY 11973, USA}
}
\\[3mm]
Wenbin Zhao
\address{Nuclear Science Division, Lawrence Berkeley National Laboratory, Berkeley, California 94720, USA}
\address{Physics Department, University of California, Berkeley, California 94720, USA}
}
\maketitle
\begin{abstract}
We review recent theoretical progress in describing collective effects in photon+nucleus collisions. The approaches considered range from the color glass condensate where correlations are encoded in the initial state, to hydrodynamic frameworks, where a strong final state response to the initial geometry of the collision is the key ingredient to generate momentum-space correlations.
\end{abstract}
  
\section{Introduction}
In high energy heavy-ion collisions, such as those performed at the Relativistic heavy-ion Collider (RHIC) and the Large Hadron Collider (LHC), the produced matter was shown to behave like an almost perfect fluid \cite{Gale:2013da}. The main experimental indication that lead to this discovery was the observation of anisotropic flow, often measured in terms of azimuthally dependent two particle correlations. In heavy-ion collisions, the interaction region is generally anisotropic in the transverse plane - for mid-central collisions, one can think of an almond shape interaction region being formed by the overlapping approximately spherical nuclei. The matter, initially contained within this anisotropic region expands and is eventually converted to individual particles that fly into the detector. Their momentum distribution is found to correlate with the expected initial shape, which requires strong interactions during the expansion stage. Hydrodynamics has been very successful in quantitatively describing the final state momentum anisotropies and multi-particle correlation observables used to probe them. In general, fluctuations of nucleon positions and subnucleonic degrees of freedom lead to fluctuating initial shapes, often characterized by a Fourier expansion in eccentricities, $\varepsilon_n$, which are the coefficients of the $e^{i n (\phi - \pi/n)}$ terms in the expansion of the spatial distribution (typically weighted by the initial energy density distribution). In analogy, final state momentum anisotropies are characterized by the Fourier coefficients of the expansion of the transverse momentum distributions, dubbed $v_n$. 

When the LHC began taking data in 2001, first results from high multiplicity p+p collisions showed long range correlations in rapidity with significant values for $v_2$ \cite{CMS:2010ifv}. This was unexpected, as such a small system was believed not to exhibit a hydrodynamic phase or any kind of significant enough final state interactions. Later results from p+Pb collisions showed even larger $v_2$ as well as higher harmonics \cite{CMS:2012qk,ALICE:2012eyl,ATLAS:2012cix}.

Both hydrodynamic model calculations as well as alternative explanations for these correlations in small systems appeared in the literature. Generally, hydrodynamic models have been rather successful in describing particle production, including particle spectra as functions of transverse momenta and flow harmonics $v_n$, in small system collisions \cite{Dusling:2015gta,Noronha:2024dtq}. 

The dominant alternative to strong final state effects is the color glass condensate calculation of multi-particle production, which predicted long range correlations with azimuthal anisotropies \cite{Krasnitz:2002ng,Gelis:2008ad,Dumitru:2008wn}. The first calculations that compared to the LHC data used the glasma graph approximation, which limit the interactions to maximally two-gluon exchanges \cite{Gelis:2008ad,Gelis:2008sz,Dumitru:2008wn,Dumitru:2010mv,Dusling:2012wy,Dusling:2013qoz}. Other calculations resummed multi-gluon exchanges and yet others treated the problem fully numerically, where multi-gluon exchanges can be included and any color charge statistics and realistic spatial distributions can be used. For detailed references see \cite{Schenke:2021mxx}. 

Generally, the CGC calculations alone have had trouble reproducing the correct multiplicity and system size dependence of the measured azimuthal momentum anisotropy. Strong final state interactions seem to be necessary to explain the experimental data even qualitatively \cite{Noronha:2024dtq}. 

This article is focused on the recent observation from the ATLAS Collaboration at LHC that even in ultra-peripheral heavy-ion collisions, which can be understood as photon+nucleus collisions, long-range correlations with azimuthal anisotropy emerge \cite{ATLAS:2021jhn}. The question is whether these collisions create a system where final state effects are less important, or whether they are very similar to proton+nucleus collisions. Given that produced particle multiplicities are similar to p+Pb for the events considered, it would not be too surprising if a $\gamma^*$+Pb collisions behaved similar to a $p$+Pb collision. In the following we will introduce ultra-peripheral collisions and sketch two distinct approaches to computing azimuthal anisotropies in $\gamma^*$+Pb collisions.

\section{Ultra-peripheral heavy-ion collisions}
Ultra-peripheral heavy-ion collisions are those where two heavy ions encounter each other at an impact parameter $|b_T|>2 R_A$, where $R_A$ is the nuclear radius. The interaction now occurs between a quasi-real photon ($Q^2 \lesssim 1/R_A^2$)from the Weizsaecker-Williams photon field of one nucleus and the other nucleus (with a nucleon, or parton, or gluon field, ... of the nucleus, depending on the kinematics). Hadronic interactions do not occur, as they are short range. Also $\gamma+\gamma$ interactions are possible, but we do not consider them in this work. 

In the high multiplicity events we consider, due to rare fluctuations with sufficiently long lifetime (longer than the time of the interaction with the nucleus), the incoming low-$Q^2$ photon can be viewed as a vector meson with a large number of collinear partons. This picture will be used for both the CGC and hydrodynamic model calculations discussed below. 

\section{Color glass condensate}
The calculation presented in \cite{Shi:2020djm}, which was the first to address azimuthal anisotropies in UPCs, models the distribution of partons in the incoming photon as a Gaussian in both transverse position and transverse momentum
\begin{equation}
   w(x,b_\perp,k_\perp) = f_{p/\gamma}(x)\frac{1}{\pi^2}e^{-b_{\perp}^2/B_p-k_\perp^2/\Delta^2}\,,
\end{equation}
where $B_p$ is the spread of partons in transverse coordinate space and $\Delta$ the typical transverse momentum of the parton. The function $f_{p/\gamma}(x)$ is the usual collinear photon parton distribution function (PDF).
The calculation uses the dilute-dense picture, which assumes a much higher parton density in the target (Pb) than in the projectile ($\gamma^*$). Multiple scattering of projectile partons with the dense target gluon fields are described using Wilson lines $U(x_\perp)$ in the eikonal approximation.
More explicitly, the production of two initially uncorrelated quarks in the dense gluon background of the target is given by
\begin{align}
    \frac{dN}{d^2k_{1\perp}d^2k_{2\perp}}=&\int_{b_1, b_2, r_1, r_2} e^{i(k_1r_1+k_2r_2)} w_1 w_2 \notag \\  &~~~~\times \left\langle D\left(b_1+\frac{r_1}{2},b_1-\frac{r_1}{2}\right)D\left(b_2+\frac{r_2}{2},b_2-\frac{r_2}{2}\right)\right\rangle\,,
\end{align}
where $D(x_\perp,y_\perp) = \frac{1}{N_c} {\rm Tr}[U(x_\perp) U^\dag(y_\perp)]$. This is the dipole operator and $x_\perp$ is the transverse position of the quark in the amplitude, and $y_\perp$ the position of the quark in the complex conjugate amplitude. Technically, this description is good for the forward (photon-going) direction, where the two quarks are going (as we probe small $x$ in the target), but the measurement is done at midrapidity. We will get back to this point when discussing another, updated CGC calculation.
The correlations we are interested in appear as higher order $N_c$ corrections in the background average of two dipole amplitudes
\begin{align}
\Big\langle D\left(b_1+\frac{r_1}{2},b_1-\frac{r_1}{2}\right)  & D\left(b_2+\frac{r_2}{2},b_2-\frac{r_2}{2}\right)\Big\rangle\Big|_{\rm up~to~\frac{1}{N_c^2}} \notag \\ &= e^{-\frac{Q_s^2}{4}(r_1^2+r_2^2)}\left[1+\frac{1}{N_c^2}Q(r_1,b_1,r_2,b_2)\right]\,,
\end{align}
where $Q(r_1,b_1,r_2,b_2)$ is the quadrupole operator, which is computed in the GBW approximation \cite{Golec-Biernat:1998zce}.

Finally, multiparticle spectra and correlations in high energy $\gamma+A$ collisions can be obtained from the Fourier transform of above dipole amplitudes, as was done for p+A collisions \cite{Mace:2018vwq}.
In UPCs the virtuality is approximately $Q\sim 30\,{\rm MeV} \ll \Lambda_{\rm QCD}$.
However, the extent of the QCD fluctuation usually does not exceed the size $1/\Lambda_{\rm QCD}$ due to color confinement, so in the calculation \cite{Shi:2020djm} $B_p= 25\,{\rm GeV}^{-2}$ was used. Further assuming $Q_s^2 = 5\,{\rm GeV}^2$, good agreement for $v_2(p_\perp)$ compared to the ATLAS data was found, at least in the region $p_\perp <2\,{\rm GeV}$. As similar calculations fail to describe the systematics with system size or multiplicity in other small systems \cite{Mace:2018vwq}, one may ask how robust this calculation is for the case of ultraperipheral collisions. 

An improvement of the CGC calculation was presented in \cite{Duan:2022pma}. By dressing the valence quarks in the projectile with gluons, it allows to compute mid-rapidity particle production involving those gluons, and it includes correlations emerging in both the projectile and target. Further, the uncertainty from the wave function of the nearly real photon is evaluated by studying two different models: the dilute quark-antiquark dipole approximation and the vector meson. The calculation is done in the so called factorized dipole approximation (FDA) \cite{Kovner:2017ssr}, because the usual approximation, the large $N_c$ limit, should not be taken, as the correlations of interest are $N_c$ suppressed. 

In the FDA, one finds the contributions that are enhanced by the area $S_\perp$, namely contributions that fulfill $Q_s^2 S_\perp\gg 1$. It turns out that the leading result in this approximation can be expressed entirely in terms of dipole operators, which simplifies the calculation. 

The results for the $p_\perp$ dependent elliptic azimuthal anisotropy differ from the previously discussed calculation qualitatively as they decrease at large $p_\perp$. The turnover of $v_2(p_\perp)$ comes from the dominance of a narrow gluon Hanbury-Brown--Twiss (HBT) peak. 

A feature common to all CGC calculation is the quick decorrealtion in $p_T$, meaning that $v_2$ drops quickly when the difference between the $p_T$ of the trigger particles to that of the associate particles is increased. Note that experimentally one defines 
\begin{equation}
    v_n(p_T^a) = v_{n,n}{(p_T^a,p_T^b)}/\sqrt{v_{n,n}(p_T^b,p_T^b)}\,,
\end{equation}
where $v_{n,n}$ are the Fourier coefficients in the expansion of the two-particle distribution
\begin{equation}
    \frac{dN}{d\vec{q}_1^2d\vec{q}_2^2} \propto 1+ 2\sum_n v_{n,n} \cos(n\Delta \phi)\,.
\end{equation}
When using extended momentum bins for $\vec{q}_1$ and $\vec{q}_2$ as done in the experiments, the fast decorrelation is smeared out, and results are compatible with experimental data, although no tuning was attempted in \cite{Duan:2022pma} to achieve a good fit.

\section{Hydrodynamics}
As hydrodynamic model frameworks have been applied to systems as small as p+p collisions, high-multiplicity $\gamma^*$+Pb collisions should not be any different in principle. Multiplicities comparable to  $p$+Pb events imply that an extended medium is formed for which final state interactions could be significant. Major differences to the hadronic collisions are the fluctuating, and significantly smaller, center of mass energy of the $\gamma^*$+Pb system, as well as the fluctuating center of mass rapidity. Further, but not as dramatic a change is the geometry of the projectile, which is assumed to be a vector meson, which is assigned one less hot spot compared to a proton projectile. The vector-meson-nucleon cross section is a major uncertainty, and assumed to equal the nucleon-nucleon cross section in this study.

This system has been modeled with 3+1D hydrodynamics, where the inclusion of longitudinal dynamics is an essential ingredient, as the particle production in this system is far from boost invariance \cite{Zhao:2022ayk,Shen:2022daw}. 
The strategy of \cite{Zhao:2022ayk} was to fit the model parameters to $p$+Pb collision data from the LHC and predict $\gamma^*$+Pb data from that constrained model. Then, successful description of the experimental data would imply that the system created in $\gamma^*$+Pb collisions behaves qualitatively similarly to that in $p$+Pb collisions.  

Indeed, it was found that the elliptic flow measured by the ATLAS Collaboration could be well described by the model. The experimental value is smaller than that in $p$+Pb collisions, which in the model results from the increased longitudinal decorrelation of the initial transverse geometry compared to the $p$+Pb case. The next higher harmonic $v_3$ is underestimated by the hydro model, yet the experimental error bars are large. The CGC models do not provide any prediction for $v_3$ in ultra-peripheral Pb+Pb collisions.

\begin{figure}[htb]
\centerline{%
\includegraphics[width=8.5cm]{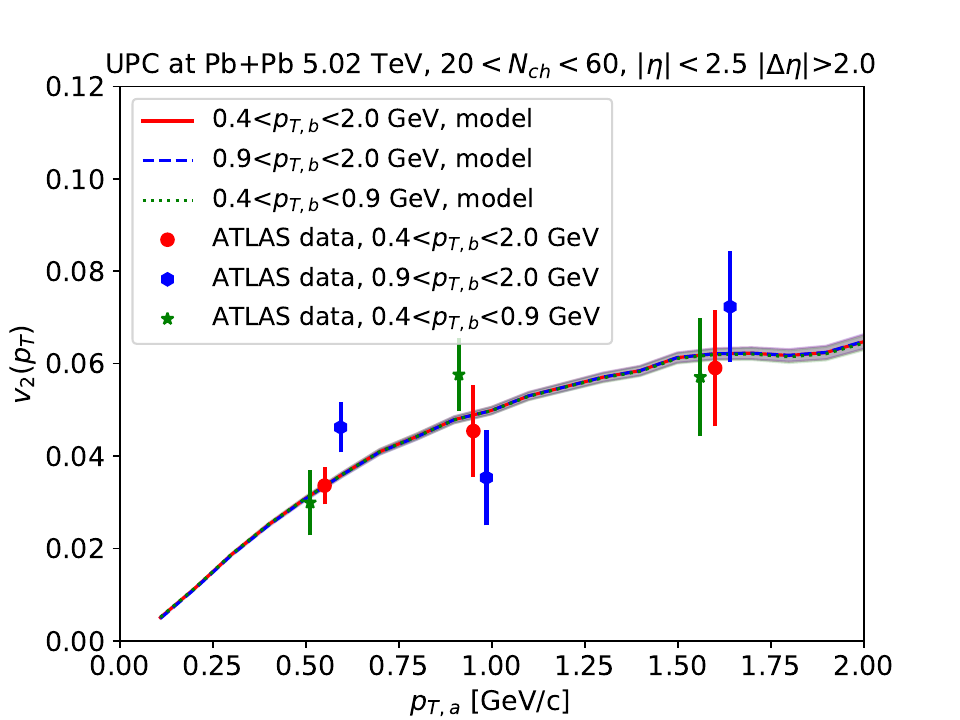}}
\caption{Elliptic anisotropy $v_2(p_\perp)$ from 2-particle correlations in Pb+Pb UPCs using different reference $p_\perp$ bins. Hydrodynamic calculations \cite{Zhao:2022ayk} compared to ATLAS data \cite{ATLAS:2021jhn}.}
\label{fig:pTdecor}
\end{figure}

A major difference to CGC calculation discussed above is that the an\-isotropy coefficients depend negligibly on the $p_\perp$ bins used to compute them. This is demonstrated in Fig.\,\ref{fig:pTdecor}, where for the hydrodynamic model, no significant dependence on the $p_\perp$ reference bin can be seen. 

Another major difference is the dependence of $v_2$ on the transverse size of the projectile. In the hydrodynamic calculation, $v_2$ increases with the transverse size, which can be explained by the increased geometric fluctuations made possible by the larger area, which lead to larger eccentricities. In the CGC, larger area leads to a larger number of independent color domains, which means particles are produced from uncorrelated regions, hence are less correlated. This opposite behavior could be studied at the future Electron Ion Collider (EIC), where the virtuality $Q^2$ dependence could be used as a proxy for the inverse transverse extent.

\section{Conclusions}
Strong final state effects have been observed in smaller and smaller collision systems. It is natural to ask whether even photon-nucleus collisions can produced hot and dense enough final states that exhibit collective behavior. While indications are there that this is indeed the case, other possible explanations of the measured $v_2$ are not fully ruled out. Further exploration of high multiplicity ultra-peripheral collisions, as well as similar events at the future EIC will help improve our understanding of collective effects in $\gamma$+nucleus collisions.

\section*{Acknowledgments}
This material is based upon work supported by the U.S. Department of Energy, Office of Science, Office of Nuclear Physics, under DOE Contract No.~DE-SC0012704 (B.P.S.), DE-AC02-05CH11231 (W.B.Z.), and Award No.~DE-SC0021969 (C.S.), and within the framework of the Saturated Glue (SURGE) Topical Theory Collaboration. C.S. acknowledges a DOE Office of Science Early Career Award. W.B.Z. is also supported by NSF under Grant No. OAC-2004571 within the X-SCAPE Collaboration, and within the framework of the SURGE Topical Theory Collaboration. 

\bibliographystyle{unsrt}
\bibliography{refs}

\end{document}